\documentclass[a4paper]{article}
\usepackage{array,booktabs}
\usepackage{geometry}
\usepackage{amsmath,amssymb}
\usepackage{graphicx}
\usepackage{color,xcolor}
\usepackage{amsfonts,amssymb}
\usepackage{longtable}
\usepackage{subfigure}
\usepackage[english]{babel}
\usepackage[utf8]{inputenc}
\usepackage{amsmath}
\usepackage{graphicx}
\usepackage{natbib}
\usepackage[colorinlistoftodos]{todonotes}
\usepackage{colortbl}
\usepackage{float}
\usepackage{epstopdf}
\usepackage{changes}

\title{Clustering Brain Signals: A Robust Approach Using Functional Data Ranking}
\makeatletter
\let\@fnsymbol\@arabic
\makeatother
\author{Tianbo Chen$^1$; Ying Sun$^1$; Carolina Euan$^1$; Hernando Ombao\thanks{Statistics Program, King Abdullah University of Science and Technology (KAUST), Thuwal 23955, Saudi Arabia. E-mails: tianbo.chen@kaust.edu.sa; ying.sun@kaust.edu.sa; carolina.euan@kaust.edu.sa; hernando.ombao@kaust.edu.sa.} }
\date{}

\begin{document}
\maketitle

\vspace{-0.5cm}
\noindent
{\bf Abstract:} In this paper, we analyze electroencephalograms (EEG) which are recordings of
brain electrical activity. We develop new clustering methods for identifying synchronized brain
regions, where the EEGs show similar oscillations or waveforms according to their spectral
densities. We treat the estimated spectral densities from many epochs or trials as functional
data and develop clustering algorithms based on functional data ranking. The two proposed
clustering algorithms use different dissimilarity measures: distance of the functional medians
and the area of the central region. The performance of the proposed algorithms is examined by
simulation studies. We show that, when contaminations are present, the proposed methods for
clustering spectral densities are more robust than the mean-based methods. The developed methods
are applied to two stages of resting state EEG data from a male college student, corresponding
to early exploration of functional connectivity in the human brain.

\newpage

\section{Introduction}
Most of the research on clustering of brain signals currently focuses on how populations of neurons respond to external stimuli or how they behave during the resting state.
Brain activity following the presentation of a stimulus and even during resting state is the result of highly coordinated responses of large numbers of neurons both locally (within each region) and globally (across different brain regions) \citep{fingelkurts2005functional}. The electroencephalogram (EEG) is a tool for monitoring the spontaneous electrical activity of the brain over a period of time. EEGs are typically recorded from multiple electrodes placed on the scalp, referred as EEG channels.
In practice, EEGs are often used to diagnose brain disorders, such as tumors, stroke, and coma, because the signals capture macroscopic oscillations caused by coordinated activities in the brain. Although the EEG has limited spatial resolution compared to functional magnetic resonance imaging, it remains a valuable tool due to its millisecond-range temporal resolution.
The goal of this paper is to develop robust time series clustering algorithms that are resistant to outliers for the identification of similar EEG channels. The clustered EEG channels are useful for understanding the functional connectivity of the brain.

Medical research and diagnostic applications generally focus on neural oscillations that are captured in EEG signals, or the spectral aspect of EEG. In this paper, we analyze the EEG data set in \cite{wu2014resting} which is a collection of dense-array EEG data from the healthy young subjects, to investigate how measures of cortical network function acquired at rest can be used to predict subsequent acquisition of a new motor skill. In statistics, various time-series clustering methods have been developed to understand the functional connectivity of the brain in the frequency domain. \cite{kakizawa1998discrimination} proposed a clustering procedure based on the Kullback-Leibler and Chernoff \citep{shumway2016time} information measures, to identify spectral features (specific frequencies and spectral density matrices). \cite{rutkowski2010emd}  decompose each of the recorded EEG channels into intrinsic mode functions (IMF) and the IMF components are further clustered for their spectral similarity in order to identify only those carrying responses to present stimuli to the subjects. \cite{Orhan201113475} decomposed EEG signals into frequency sub-bands using a discrete wavelet transform, and the wavelet coefficients were clustered using the K-means algorithm for each frequency sub-band. \cite{Maharaj12} applied wavelet methods for multivariate time-series clustering and evaluated the performance of different clustering methods for stationary and variance nonstationary multivariate time series with different error correlation structures. \cite{purdon2013electroencephalogram} proposed the use of coherence across the entire range of frequencies to analyze the spatiotemporal dynamics of multivariate time series. \cite{JTSA:JTSA12166} developed a new method model for multiple groups of time series, accounting for both between-group and within-group differences in EEG signals spectra. \cite{euan2016hierarchical} proposed the spectral merge clustering algorithm to cluster EEG channels with similar spectra to identify the EEG channels that share similar spectral features. The similarity was measured by total variation distance to measure the similarity among EEG spectra from different channels. There are also time-series clustering methods that are not in the frequency domain. \cite{panuccio2002hidden} presented the Hidden Markov models for the online classification of the single epoch EEG data during imagination of a left or right-hand movement.

EEGs are a non-invasive way of indirectly measuring neuronal electrical activity.
The key challenge is that EEGs are often noisy and have outliers which makes the statistical modeling and inference more challenging. However, to the best of our knowledge, robust methods for EEG analysis are sparse when outliers and contaminations are present.  \cite{ngo2015exploratory}  proposed the use of functional boxplots \citep{sun2012adjusted} on log-periodograms to visualize and analyze EEG data, where the functional median and the variability of the smoothed log-periodograms are summarized by the functional boxplots, and potential outliers are also detected. By applying functional boxplot, we can analyze EEG data with outliers and noise. \cite{hasenstab2016robust} proposed a robust functional clustering (RFC) algorithm to identify subgroups within EEG data. The RFC is a model-based algorithm based on functional principal component
analysis.

In this paper, we develop robust clustering algorithms to identify synchronized brain regions that show similar oscillatory patterns. We propose two robust clustering algorithms that are both based on functional data ranking but using different dissimilarity measures: the distance of functional medians and the area of the central region. Finally, we visualize different clustering results for early and late stages of the experiment to explore the functional connectivity of the human brain. We point out that both the EEG data analysis and the proposed algorithms are novel. In the EEG data analysis, we segment each EEG recording into multiple time series per second, so that the robust functional analysis methods can be used to eliminate the potential contaminations during the one-second time interval. The segmentation also ensures the stationarity in the one-second epoch due to the resting-state testing and the short-time epoch \citep{euan2016hierarchical}. For clustering algorithms, the two proposed similarity measures are also innovative. Compared to the mean-based clustering methods, they are robust and resistant to outliers. Moreover, the central region method measures the similarity of the majority of two clusters taking the uncertainty into account, which is a non-trivial task for clustering functional data.

The rest of the paper is organized as follows. The background for spectral analysis and functional boxplot are introduced in Section 2, and the robust clustering algorithms are proposed in Section 3. In Section 4, we investigate the performance of different clustering algorithms by simulation studies. The application to the EEG is presented in Section 5. In Section 6, we conclude the paper.

\section{EEG Data and the Spectral Density Estimation}

In this section, we first describe the EEG data that will be used in Section 2.1 and then introduce the spectral analysis that how to estimate the spectra in Section 2.2

\subsection{EEG Data Description}

\begin{figure}[H]
\centering
\raisebox{-5cm}{\includegraphics[width=11cm]{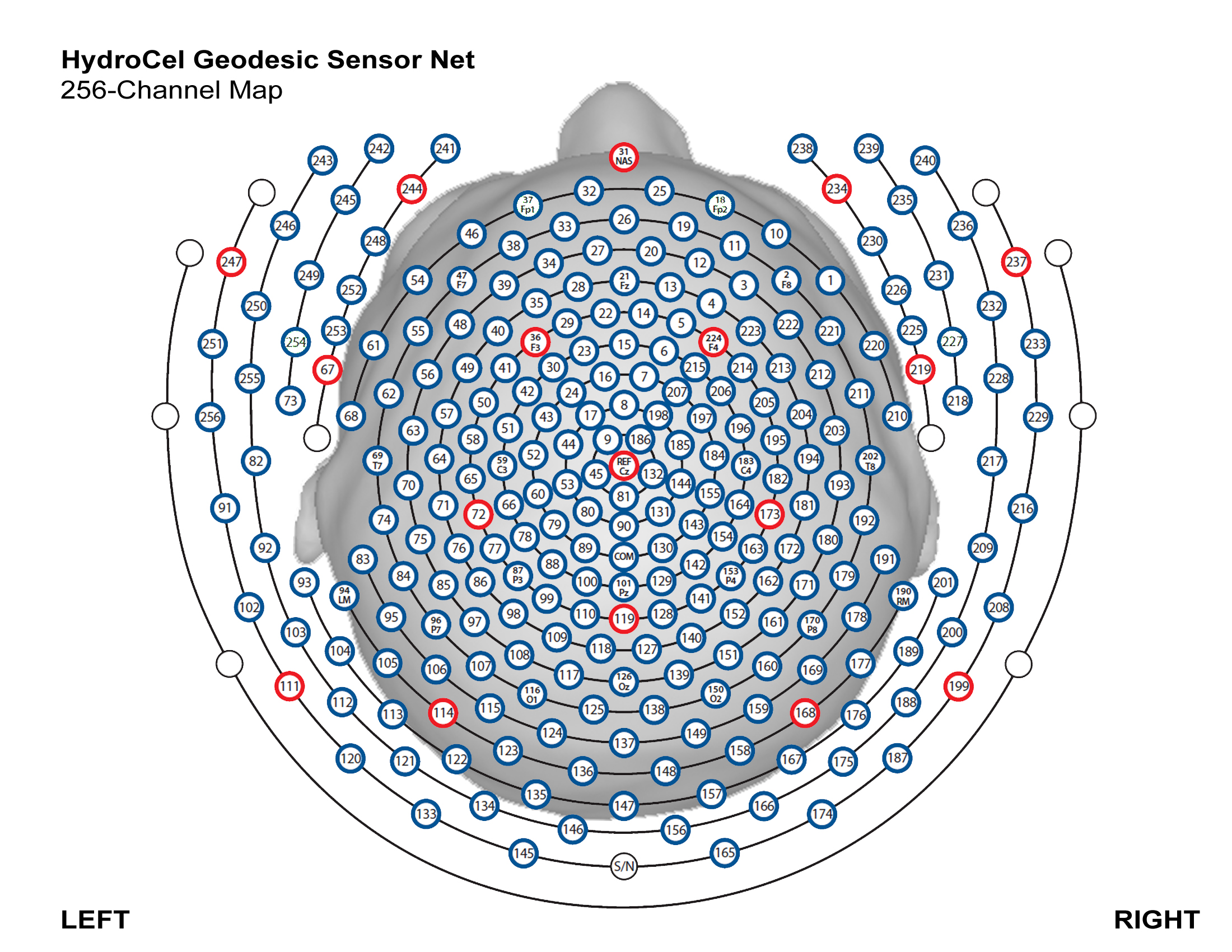}}
\caption{The locations of the $256$ channels on the scalp surface.}
\label{256}
\end{figure}

The EEG data was collected
from an experiment described in \cite{wu2014resting}.
EEG signals were recorded from $256$ channels on the scalp that are shown in Figure \ref{256}, with a millisecond resolution ($1000$ recordings per second). From the $256$ channels, $62$ channels were eliminated due to  issues with data quality (in particular, large artifacts).
The total recording time for the EEG data is $177$ seconds. Here, we will focus on investigating the brain connectivity during the first minute and the last minute of a resting state. We treat the recordings of each second as the unit of a time series to create $60$ epochs (trials) of data for a given minute.

\subsection {Spectral Analysis}
Let ${\bf x}=\{x_t\}$, $t=1,...,l$, denote a zero-mean weakly stationary EEG time series in an epoch, and let $\gamma(h)$ , denote its autocovariance function that satisfies
$\sum_{h=-\infty}^{\infty}|\gamma(h)|<\infty$, where $h$ is the time lag.
Then $\gamma(h)$ has the following representation:
$\gamma (h)=\int_{-1/2}^{1/2}$ exp$(2\pi i \omega h)f(\omega)d\omega, \text{  } \text{  } \text{  } h=0,\pm1,\ldots, $
where $f(\omega)$ is the spectral density of  ${\bf x}$, satisfies
$$f(\omega)=\sum_{h=-\infty}^{\infty}\gamma(h)e^{-2 \pi i \omega h}, \text{  } \text{  } -1/2\leq \omega \leq 1/2.$$

A periodogram is a sample version  of the spectral density that can be used to estimate the spectral density function. \cite{caiado2006periodogram} proposed to consider periodogram as a feature for the purpose of clustering many time series. The periodogram of ${\bf x}$ is calculated by $I(\omega_j)=|d(\omega_j)|^2,$
where $d(\omega_j)$ is the {\bf discrete Fourier transform (DFT)}:
$$d(\omega_j)=l^{-1/2}\sum_{t=1}^lx_te^{-2\pi i \omega_j t},\text{  }\text{  }j=0,1,...,l-1.$$
The frequencies $ \omega_j= j/l$ are called the Fourier or fundamental frequencies \citep{shumway2016time}.

In resting-state EEG data analysis, researchers are more interested in the low-frequency band since the high-frequency band does not contribute significantly to the total power of the EEG data. In the EEG data analysis in this paper, we consider the first $T=50$ frequencies: $\omega=\{\omega_0,...,\omega_{49}\}$, which correspond to $0$ - $49$ Hertz. Raw periodogram curves are often very noisy, in order to produce good clustering results, we smooth the periodograms before clustering. Smoothed periodograms are mean-squared consistent estimators, and in this paper, the smoothing bandwidth is selected automatically using the gamma-GCV method in \cite{ombao2001simple}. Moreover, while the periodogram is an approximately unbiased estimator of the spectrum, the log-periodogram is no longer (approximately) unbiased for the log spectrum due to Jensen’s inequality. Thus, we shall use the bias-corrected log periodogram $y(\omega_j)=$log$\{I(\omega_j)\}+\gamma$, where $\gamma=0.57721$ is the Euler Mascheroni constant (\cite{wahba1980automatic}, \cite{freyermuth2010tree}).

\section{Algorithms for Clustering EEG Signals}
For the EEG data of a single subject, we observe the EEG signal from $m$ channels, and each channel contains $n$ trials (epochs). Let ${\bf y}_{ik}(\omega)$ be the smoothed log-periodogram from the $i$-th channel and $k$-th epoch, where $i=1,...,m,\text{ }k=1,...,n,$. Let ${\bf y}_i^*(\omega)$ be the functional median of all the $n$ epochs in the $i$-th channel. In our application, $m=194$ and $n=60$. We propose two hierarchical clustering algorithms with different similarity measures. The similarity measures are based on the concept of functional median and central region from the tool, functional boxplot, developed by \citep{sun2012functional}. The details of the functional boxplot are provided in the supplementary material. The functional-median-based algorithm (FM) is presented in Section 3.1, and the central-region-based algorithm (CR) is in Section 3.2.

\subsection{Functional-Median-Based Algorithm (FM)}
The FM aims to cluster channels whose functional medians are similar. The advantages of using functional median are that it is one of the observed smoothed log-periodograms, and it is robust to contaminations.
We consider the {\bf Euclidean distance} between two functional medians ${\bf y}_1^*(\omega)$ and ${\bf y}_2^*(\omega)$ which are defined to be
$$D_{El}\{{\bf y}_1^*,{\bf y}_2^*\}=\left[\sum_{j=1}^T\{{\bf y}_1^*(\omega_j)-{\bf y}_2^*(\omega_j)\}^2\right]^{1/2}.$$
Other distance measures will be shown in Section 6.2.

\subsubsection{Algorithm: FM}
Denote the smoothed log-periodograms of the $m$ EEG channels by ${\bf Y}_1,...,{\bf Y}_m$, and ${\bf y}_1^*$,...,${\bf y}_m^*$ be the corresponding functional medians. Then, we have the FM algorithm:\\
{\bf Step 1}: Set initial clusters to be ${\bf C}=\{C_1,C_2,...,C_m\}$, where $C_i={\bf Y}_i,\text{ } i=1,...,m$. Then, the initial number of clusters of clusters $N=$ length $({\bf C})=m$.\\
{\bf Step 2}: Compute the distance matrix $D=\{d_{ij}\}_{N\times N}, \mbox{where}\text{ }d_{ij}=D_{El}({\bf c}_i^{*},{\bf c}_j^{*})$, where ${\bf c}_i^{*}$ and ${\bf c}_j^{*}$ are the functional medians of $C_i$ and $C_j$, respectively.\\
{\bf Step 3}: Merge $C_p$ and $C_q$ if $d_{pq} =$min$(d_{ij})$ and  $C_{new}=C_p \cup C_q$. Then $N=N-1$ and ${\bf C}=\{{\bf C}\setminus\{C_p,C_q\},C_{new}   \}$.\\
{\bf Step 4}: Repeat Steps 2 and 3 until the number of clusters reaches the prespecified number of clusters.

\subsection{Central-Region-Based Algorithm (CR)}
The CR aims to cluster channels that are similar in the majority, taking the variability of the clusters into account. The advantages of the proposed CR algorithm are: (i.) it considers the information of the 50\% most central curves; (ii.) it takes the variability in the distribution of the sample curves into account; (iii.) it is robust to the contaminations.

\subsubsection{Similarity Measure}
We propose to use the area of the $50\%$ central region after merging two clusters to measure the similarity. If the central region is small, it implies that the majority of the two clusters is similar with small variation. Here, we show an example of three clusters to illustrate the similarity measure based on central regions. We define $CR(X,Y)$ as the area of the central region after merging clusters $X$ and $Y$.

\begin{figure}[H]
\begin{center}
\includegraphics[width=\textwidth]{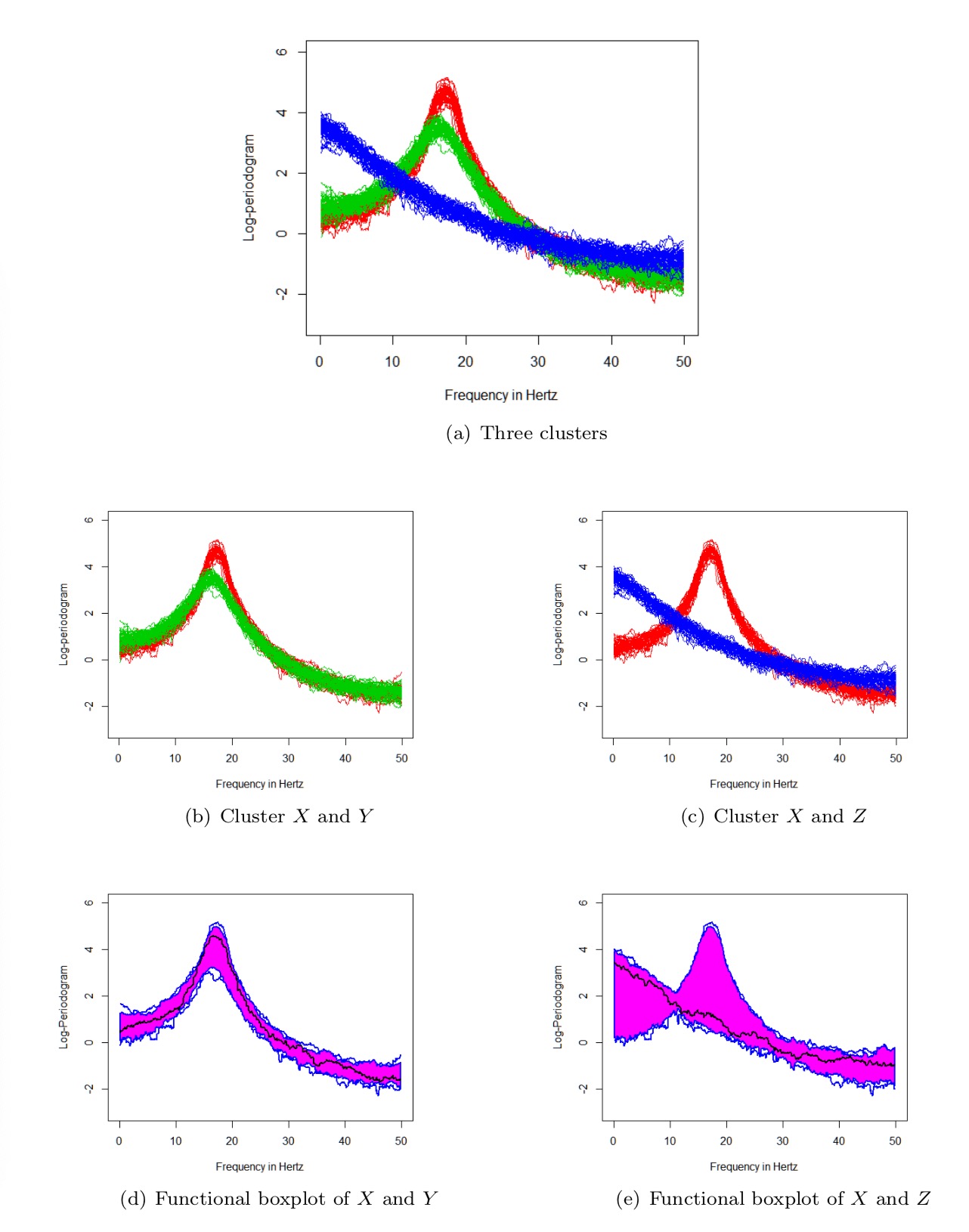}
\end{center}
  \linespread{0.3}
  \caption{(a) The three clusters, (b) the clusters $X$ and $Y$, (c) the clusters $X$ and $Z$, and (d), (e) the corresponding $50\%$ central regions after merging.}
  \label{3clu2}
\end{figure}

Figure \ref{3clu2}(a) shows the log-periodograms from the three different clusters, among which clusters $X$ (red) and $Y$ (green) are similar. Figure \ref{3clu2}(d) and (e) show the area of the $50\%$ central regions. We can see that $CR(X,Y)=475<CR(X,Z)=926$, which means that cluster $X$ is more similar to cluster $Y$ than to cluster $Z$ (blue).

\subsubsection{Algorithm: CR}

{\bf Step 1}: Set clusters to be ${\bf C}=\{C_1,C_2,...,C_m\}$ where $C_i={\bf Y}_i,\text{ } i=1,...m$. Then, the initial number of clusters of clusters $N=$ length $({\bf C})=m$.\\
{\bf Step 2}: Compute the central region matrix $S=\{s_{ij}\}_{N\times N},\text{ }s_{ij}=CR(C_i,C_j)$.\\
{\bf Step 3}: Merge $C_p$ and $C_q$ if $s_{pq} =$min$(\{s_{ij}\})$ and  $C_{new}=C_p \cup C_q$. Then $N=N-1$ and ${\bf C}=\{{\bf C}\setminus\{C_p,C_q\},C_{new}   \}.$\\
{\bf Step 4}: Repeat Steps 2 and 3 until the number of clusters reaches the prespecified number of clusters.\\

\section{Simulation Studies}
In this section, we use simulated data to test our proposed clustering algorithms. We propose to use mixture $AR(2)$ models in the simulation studies. One way to model the EEG signals to capture their oscillatory activity is via a mixture of second order auto-regressive ($AR(2)$) processes:
$x_t=\phi_1 x_{t-1}+\phi_2 x_{t-2}+w_t,$
where $w_t$ is the white-noise process with variance $\sigma_w^2$. Another way of defining an $AR(2)$ process is through the polynomial operator. Define the backshift operator $B^k x_t = x_{t-k}$, where $k$ is a non-negative integer. Then an $AR(2)$ process can be defined as
$\phi(B)x_t=w_t,$
where $\phi(z)=1-\sum_{k=1}^2\phi_kz^k$. Then the spectral density for the $AR(2)$ process is given by
$$f(\omega)=\frac{\sigma_w^2}{|\phi(e^{-2\pi i \omega})|^2}=\frac{\sigma_w^2}{| 1 - \phi_1 e^{-2 \pi i \omega} - \phi_2 e^{-4 \pi i \omega} |^2}.$$

We introduce two types of contaminations to test the robustness of the proposed algorithms and explain the performance of each algorithm. To show the robustness, we use the mean-based algorithm as the comparison. Similar to the FM algorithm, but in step 2, we compute the distance of the functional means of the two cluster, instead of the functional medians.
\vspace{-0.3cm}
\subsection{Data Generation}
We generate the dataset from mixture $AR(2)$ models \citep{gao2016evolutionary}. Define ${\bf Y}_k=\{{\bf Y}_k(l),l=1,...,T\}$ to be a latent $AR(2)$ source whose spectrum contains a peak at precise frequency bands: $k=1$ indicates a peak at the delta band (0-4 Hz), $k=2$ indicates a peak at the theta band (4-8 Hz), $k=3$ indicates a peak at the alpha band (8-16 Hz), $k=4$ indicates a peak at the beta band (16-32 Hz) and $k=5$ indicates a peak at the gamma band (32-50 Hz). These bands were also discussed in \cite{ngo2015exploratory}.

For the $5$ latent $AR(2)$ sources, we regard each latent source as a cluster and we generate $200$ time series for each cluster. Each cluster has $5$ channels and each channel has $n=40$ epochs. The total number of channels is $m=25$ and the length of the time series is $l=1000$. The $AR$ coefficients of the $5$ clusters are (0.8,0.1),(0.9,-0.9),(-0.1,-0.9),(-0.9,-0.9,) and (-0.8,-0.1). To make the channels in each cluster slightly different, we add small differences to all the $\phi_1$s of the channels in each cluster and the differences follow independent $N(0,0.01)$. These differences were small perturbations and the resulting processes turned to still be causal.

\begin{figure}
\centering
\raisebox{-2cm}{\includegraphics[width=\textwidth]{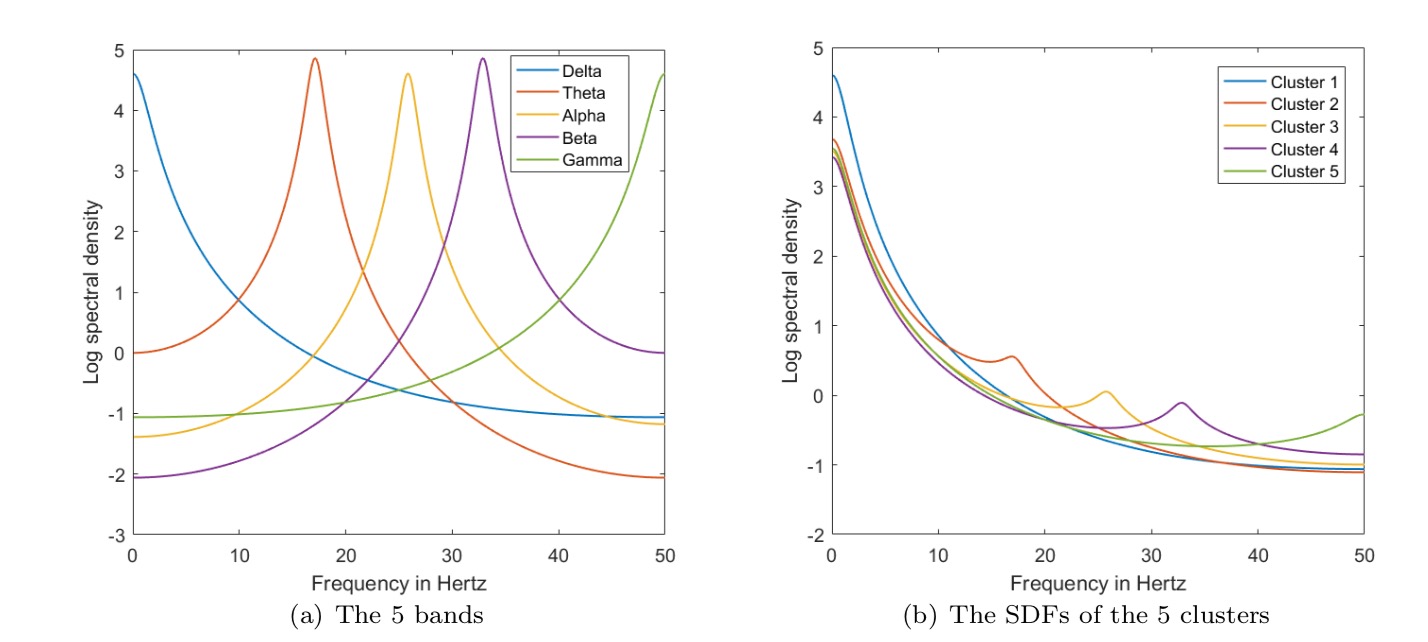}}
\linespread{0.3}
\caption{(a) Log spectral densities of the 5 bands and (b) the log spectral densities of the 5 clusters. The $AR$ coefficients of the 5 bands are (0.8,0.1), (0.9,-0.9), (-0.1,-0.9), (-0.9,-0.9) and (-0.8,0.1).}
\label{2datasets}
\end{figure}

The mixture $AR(2)$ data also have $5$ clusters, which are the linear combination of the $5$ latent sources. Suppose that ${\bf Y}=({\bf Y}_1^\top,{\bf Y}_2^\top,{\bf Y}_3^\top,{\bf Y}_4^\top,{\bf Y}_5^\top)$ denote the $5$ latent sources, and ${\bf Z}=({\bf Z}_1^\top,{\bf Z}_2^\top,{\bf Z}_3^\top,{\bf Z}_4^\top,{\bf Z}_5^\top)$ denote the $5$ clusters of the mixture data to be used in the simulation studies. The mixture $AR(2)$ data also have $m=25$ channels and $n=40$ epochs in each channel. We have ${\bf Z^\top}={\bf AY}^\top,$ where ${\bf A}$ is the weight matrix. Figure \ref{2datasets}(b) shows the true spectral densities of the $5$ clusters when ${\bf A}$ is:
$$\left[
 \begin{matrix}
   1 & 0 & 0 & 0 & 0 \\
   4/5 & 1/10 & 0 &  0  &  0 \\
   3/5 & 0 & 1/10 & 0 & 0 \\
   2/5 & 0 & 0 & 1/10 & 0 \\
   1/5 & 0 & 0 & 0 & 1/10\\
   \end{matrix}
 \right]$$

\subsection{Contaminations Models}
We add contaminations to the epochs to test the robustness of our clustering methods. Let $\delta$ be the contamination rate,
indicating that the data in each epoch have a probability of $\delta$ to be contaminated. For type 1 contamination, $\delta$=0.1, 0.2 and 0.3, and for type 2 contamination,
$\delta$=0.25, 0.3 and 0.35, corresponding to an eyeblink every 3 to 4 seconds.
We consider two types of contaminations in this simulation study. Type 1 contamination is adding a shift to
a log-periodogram \citep{sun2012functional}.  This is equivalent to adding a white-noise process to the signal. 

\begin{figure}
\centering
\raisebox{-2cm}{\includegraphics[width=0.9\textwidth]{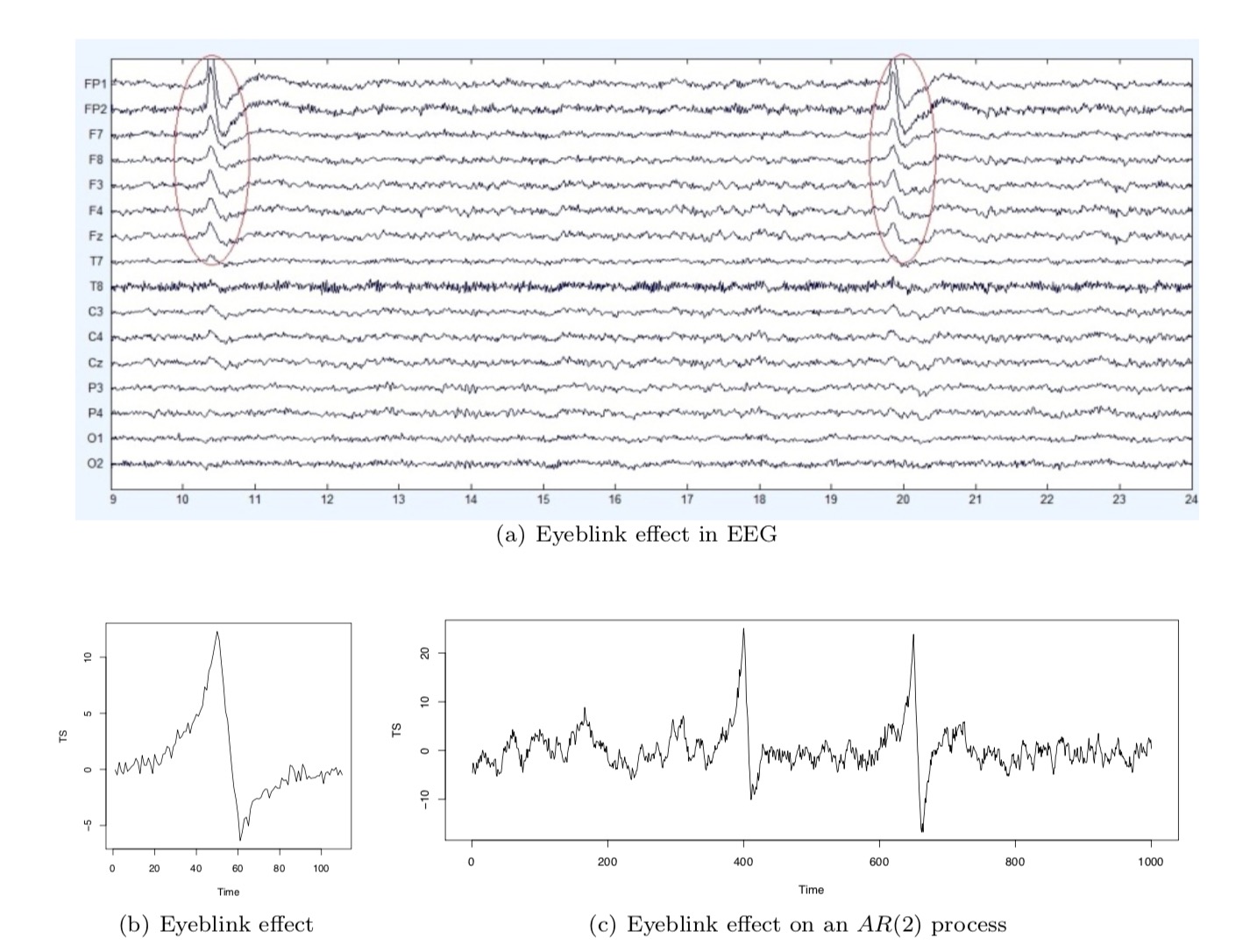}}
\linespread{0.3}
\caption{(a) The eyeblink effect in EEG data and (b), (c) the generated eyeblink effect.}
\label{eye2}
\end{figure}

For type 2 contamination, we consider adding eyeblink effect to the time series. \cite{viqueira2013ocular} studied the ocular movement and cardiac rhythm control using EEG. During EEG testing, there can appear different signals called artifacts. One of the causes of the artifacts can be the ocular movement, such as eye blinking.
Figure \ref{eye2}(a) shows the EEG with two eyeblink effect. We use the difference of two gamma functions added with white noise to simulate the eyeblink effect. Figure \ref{eye2}(b) shows the generated eyeblink effect, and Figure \ref{eye2}(c) shows how the eyeblink effect affect an $AR(2)$ process.

\subsection{Measure of Quality}
We use adjusted Rand index \citep{nguyen2009information} in {\sf R} package {\sf TSclust} \citep{montero2014tsclust} to evaluate the clustering results. The ARI ranges from $0$ to $1$, with $0$ indicating that the two clustering results do not agree
on any pairs, and the ARI is $1$ when the two clustering results are the same.



\subsection{Comparisons}
We generate the mixture $AR(2)$ model $100$ times and randomly add different rates of contaminations. We apply the proposed clustering methods to the contaminated data and compute the ARI values for each simulation run. We also cluster the data using the mean-based algorithm. The mean based algorithm merges two clusters where their functional means have the smallest  distance. We consider two distance measures:
\begin{itemize}
\item The Euclidean distance that described in Section 3.1.
\item The {\sf diss.SPEC.LLR} distance \citep{kakizawa1998discrimination, vilar2004discriminant} in the {\sf R} package {\sf TSclust}.
\end{itemize} 
We also show the results using the total variation distance (TVD, used in \cite{euan2016hierarchical}) in the supplementary file.

Table 1 shows the averaged ARI values for each of the two contamination types. We can see that when there are no contaminations, all
the four algorithms give high Rand index values, indication high similarity comparing to the true clusters. But when contaminations are present, for all the
contamination rates, the mean-based algorithms have the lowest ARI values, indicating lowest rate of agreement to the true clusters. The two proposed methods
perform better than the mean-based algorithms. The averaged computation time for the four algorithms are 64.47 seconds, 55.57 seconds, 33.99 seconds, and 32.85 seconds, respectively.
\begin{table}


\begin{center}
\def\arraystretch{0.5}
\begin{tabular}{ccccccc}
\hline
  Contamination & $\delta$    &Measure &   FM  & CR  & Mean-Euclidean & Mean-LLR   \\\hline
  Null     & 0     & ARI    &1.0000  & 0.9869 &1.0000 & 1.0000\\\hline
  Type 1 & 0.1   & ARI &   1.0000  &0.9726 &0.7532 &  0.7549\\
  & 0.2           & ARI   &   1.0000  &0.9431 &0.6805 &0.6786\\
   & 0.3            & ARI&    0.9327  &0.8798 &0.6251 &0.6244\\\hline
  Type 2 & 0.25& ARI &   1.0000    & 0.9384 & 0.8842 & 0.8853\\
   & 0.3            & ARI&   0.9889    & 0.9367 & 0.8649  &0.8649\\
   & 0.35           & ARI &   0.9074    & 0.8413 & 0.8301  &0.8302\\\hline
\end{tabular}\\
\linespread{0.3}
\caption{The averaged ARIs of the four algorithms with different contamination types and rates. The values are computed using a personal computer with 2.6 GHz Intel Core i7-9750H and 32 GB memory. }
\end{center}
\end{table}

\begin{figure}[!ht]
\centering
\includegraphics[width=12cm]{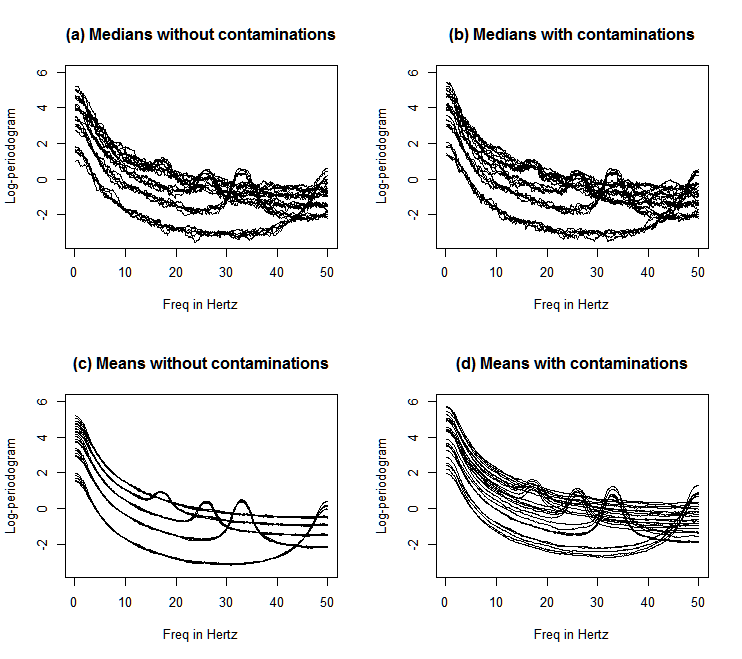}
\linespread{0.3}
\caption{(a) The functional medians of the $25$ channels and (b) the functional medians with type 1 contaminations. (c) The means of the $25$ channels and (d) the means with type 1 contaminations.}
\label{ctb}
\end{figure}

Figure \ref{ctb} illustrates the difference between the functional means and functional medians.
We plot the functional medians and functional means of the $25$ channels in the cases of no contaminations
and added with type 1 contaminations. We can see that when there are no contaminations, the mean functions
are smooth while the functional medians are noisy. However, when type 1 contaminations are added, the functional
means are shifted but the functional medians remain similar.

\begin{figure}[!ht]
\centering
\raisebox{-3cm}{\includegraphics[width=0.8\textwidth]{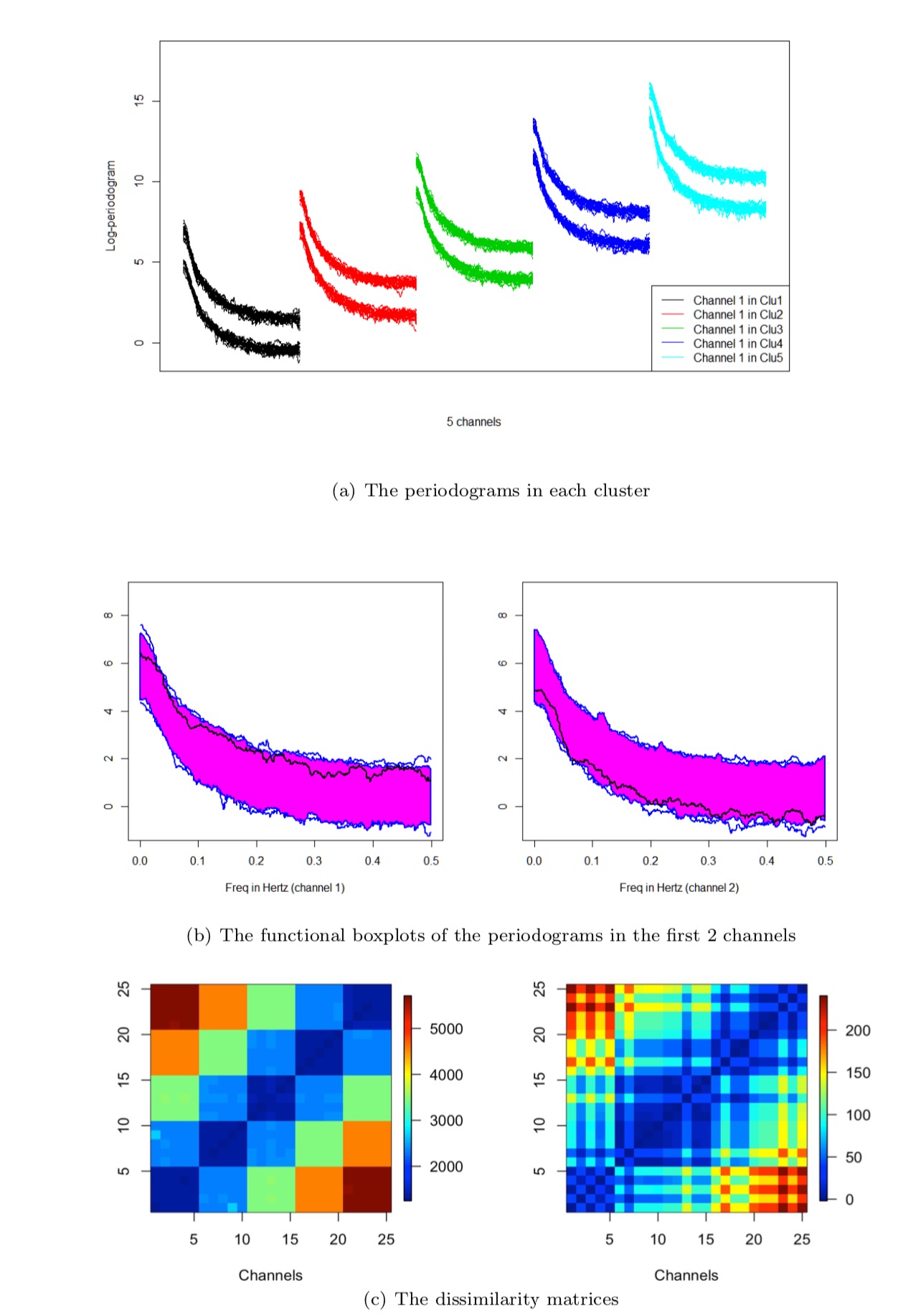}}
\linespread{0.3}
\caption{(a) The periodograms in the first channels of the 5 clusters, (b) the functional boxplots of the periodograms in the first 2 channels and (c) the dissimilarity matrices of CR (left) and FM (right).}
\label{special}
\end{figure}

Comparing the two proposed methods, FM performs better than CR.
The reason is that FM considers the most central curve and has less variability as a summary of
the centrality when the distribution of curves has high density in the center and sparse at the tails.
When curves are dense in the center, the functional median is stable and can be estimated accurately.
When the distribution is skewed or tends to be bimodal, the most central curves are more spread out
and the central region that constructed by the first 50\% central curves is clearly a better statistics to describe the centrality.  Now we consider
another simulation design with $m=5$, $n=40$ and the length of the periodograms is 500. The periodograms
are shown in Figure \ref{special}(a). We can see that in each channel, the periodograms are bimodal
and concentrated on the two side. This can cause functional medians unstable, and cannot represent
the center of the distribution for each channel. We run 100 simulations and apply type 2 contaminations, where the difference between FM and CR is larger than applying type 1 contaminations in Table 1. The results are shown in Table 2, where CR outperforms FM. This can be explained by the functional boxplots in Figure \ref{special}(b) and (c): the functional medians of the two channels from the same cluster are different while their central regions remain similar; and the dissimilarity matrices explain why FM has a higher misclassification rate.

\begin{table}[H]
\begin{center}
\def\arraystretch{0.5}
\begin{tabular}{ccccc}
\hline
  Contamination & $\delta$    &Measure &   FM  & CR     \\\hline
  Null     & 0     & ARI      &0.4802  & {\bf 1.0000} \\
  Type 2 & 0.1   & ARI &   0.4969  &{\bf 1.0000}\\
  Type 2 & 0.1   & ARI &   0.4853  &{\bf0.9989} \\
  Type 2 & 0.25& ARI &    0.4910   & {\bf 0.9979} \\\hline

\end{tabular}
\linespread{0.3}
\caption{The averaged ARIs of the FM and CR with type 2 contaminations.}
\end{center}
\end{table}

\section{Application to EEGs}
In this section, we will apply different clustering algorithms to EEGs. Here one critical issue is to identify the number of clusters which is usually unknown for real data. We propose to use the elbow method, as in \cite{euan2016hierarchical}, to choose the number of clusters in Section 5.1, the clustering results obtained from different algorithms (FM, CR, AND mean-Euclidean) are presented in Section 5.2, and the robustness test of the clustering algorithms is in Section 5.3.

\subsection{Identifying the Number of Clusters }
Before applying the algorithms to EEGs, we used the elbow method to determine the number clusters.
Recall that in the hierarchical merging algorithm, in each iteration, we merged two clusters with minimum dissimilarity (minimum distance of functional medians or minimum area of central regions) and the number of clusters was reduced by $1$. At the beginning, we merged the most similar channels (should be within a cluster), but the dissimilarity increased slowly until the optimal number of clusters was reached. If we continue, clusters will be merged, which causes the dissimilarity to increase dramatically. We plot the minimum dissimilarity against the number of clusters to find the ``elbow” in which can be used to identify the number of clusters.

\begin{figure}[!ht]
\centering
\raisebox{-2cm}{\includegraphics[width=0.75\textwidth]{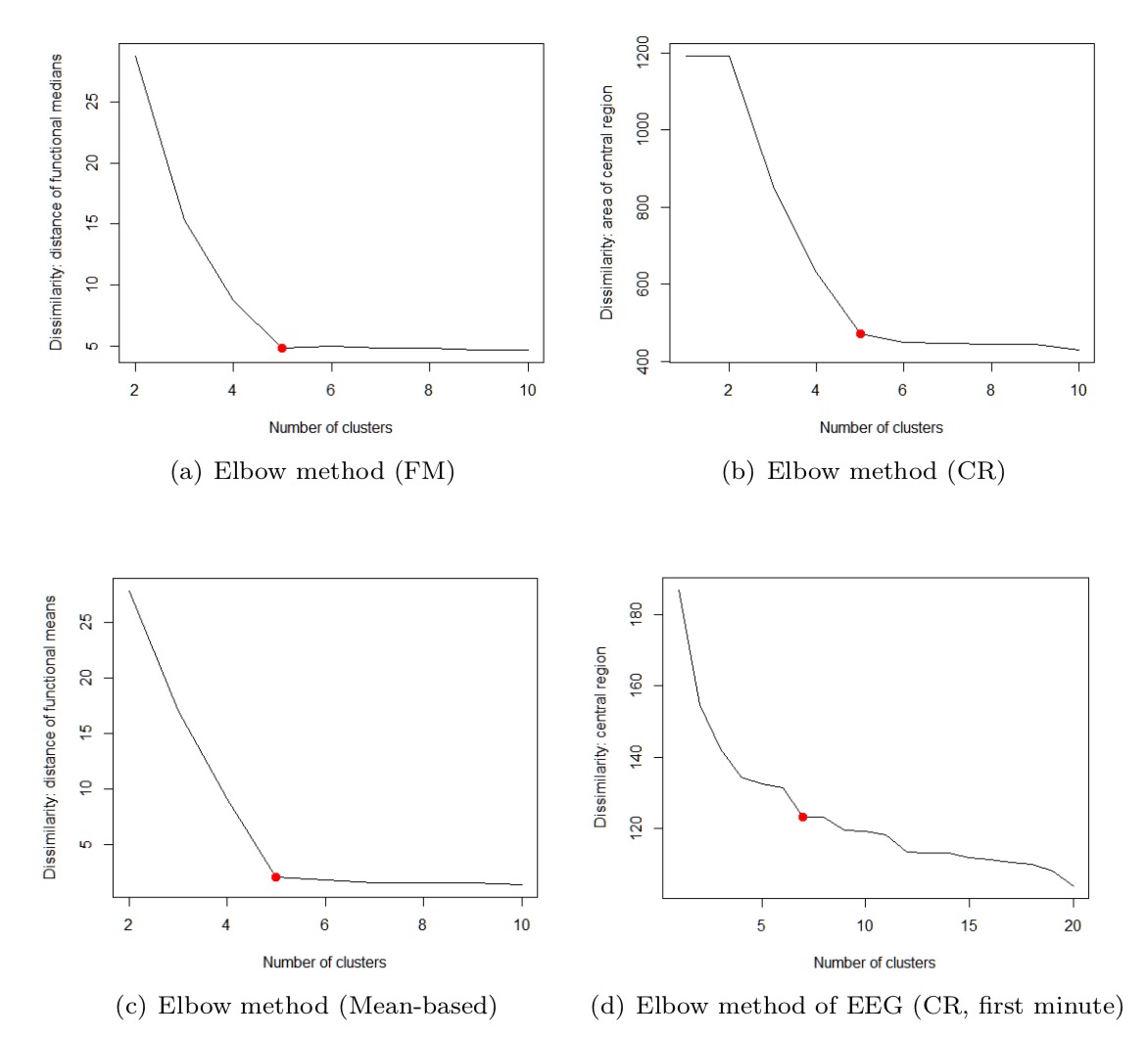}}

\linespread{0.3}
\caption{(a) The elbow method in the simulation study using FM. (b) The elbow method in the simulation study using CR, (c) the elbow method in the simulation study using functional mean algorithm and (d) the elbow method of the first minute EEG data using CR.}
\label{scree}
\end{figure}

For the simulation study in Section 4.1, where the true number of clusters is $5$, we show the three elbow method using three clustering algorithms of the first replicate data in Figure \ref{scree}(a)-(c). Note that in the final step of clustering, all the log-periodograms are clustered together. At this time we can calculate the central region area of the one cluster but we cannot calculate the distance between functional medians since there is only one functional median. So that in the elbow method of CR, the x axis begins at $1$ cluster, and in the elbow method of FM, the x axis begins at $2$ clusters.

For the EEG data of the first minute, one example of the elbow method is shown in Figure \ref{scree}(d).  For the 194 channels, we considered setting 3 as the minimum number of clusters and 20 as the maximum number of clusters. As illustrated in the elbow method, seven clusters were chosen. The number of clusters in the first minutes of the three methods (FM, CR, and mean-based) are 11, 7, and 7. The number of clusters in the last minutes of the three methods are 10, 7, and 8.

\subsection{Results}
\begin{figure}[!ht]
\centering

\raisebox{-2cm}{\includegraphics[width=\textwidth]{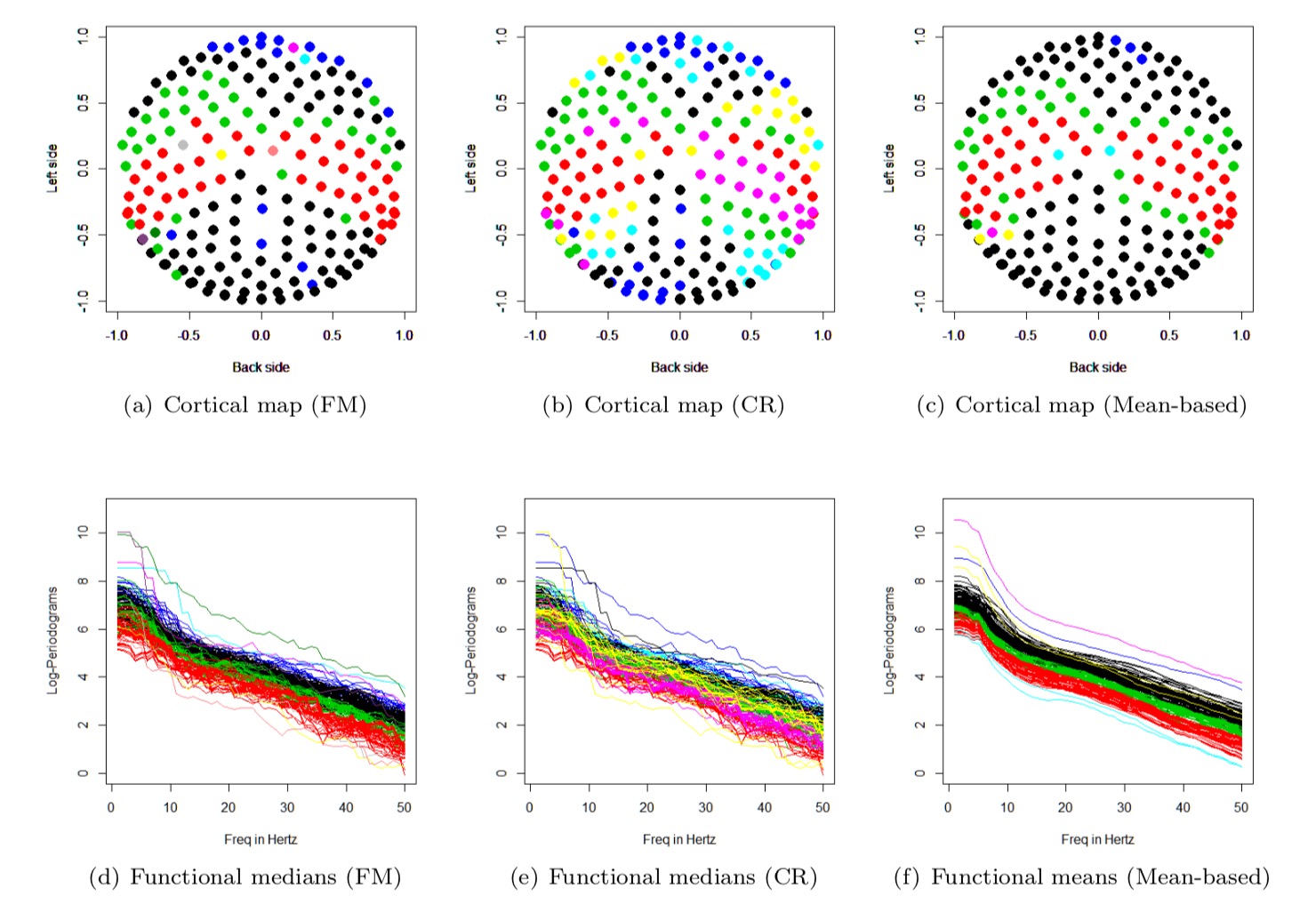}}

\linespread{0.3}
\caption{The result of the first minute. (a) The 2D cortical map of the FM, (b) the 2D cortical map of the CR, and (c) the 2D cortical map of the mean-based algorithm. (d) The functional median log-periodogram of each channel (FM), (e) the functional median log-periodogram of each channel (CR), and (f) the functional mean log-periodograms of each channel.}
\label{first}
\end{figure}

\begin{figure}[!ht]
\centering

\raisebox{-2cm}{\includegraphics[width=\textwidth]{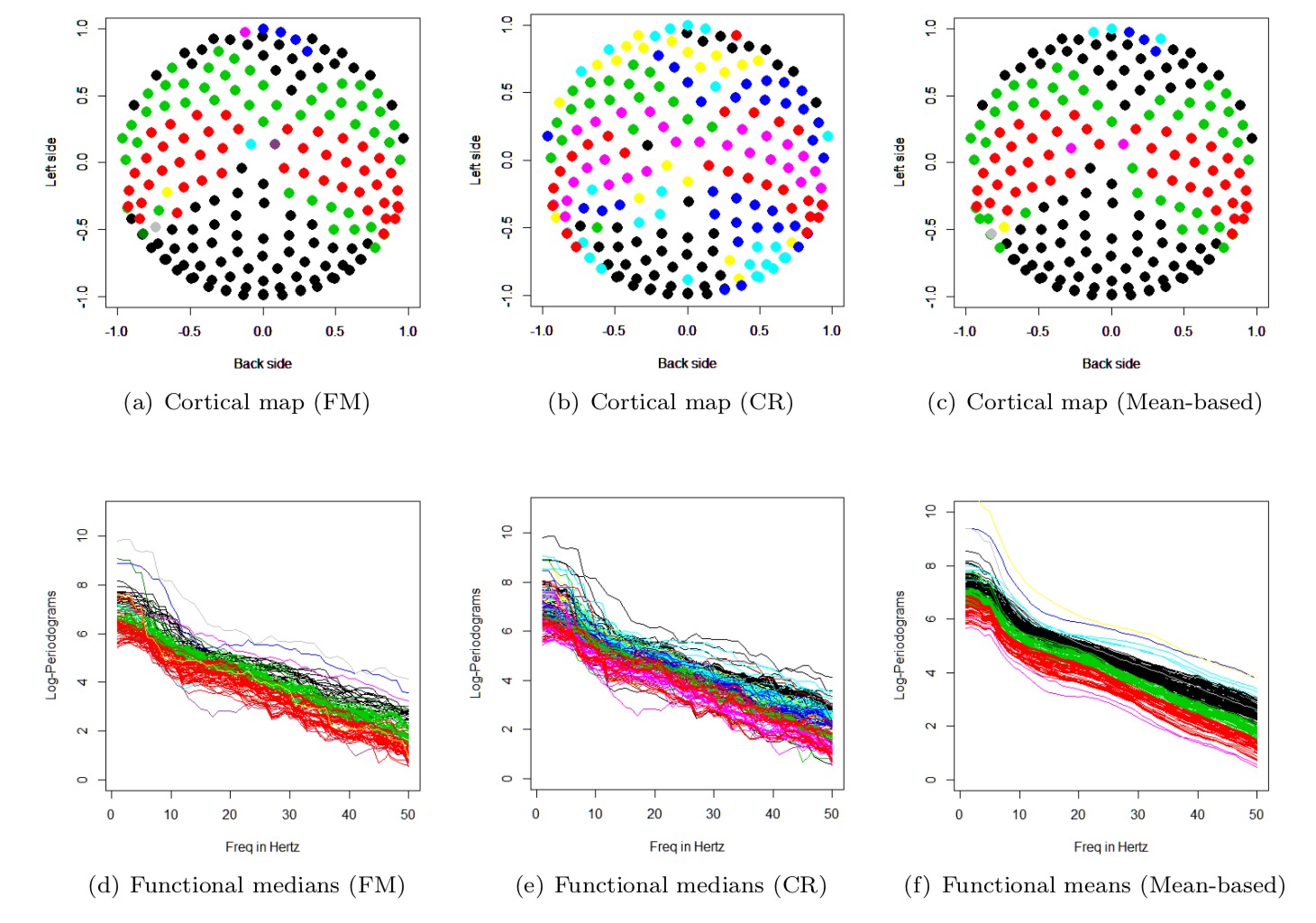}}

\linespread{0.3}
\caption{The result of the last minute. (a) The 2D cortical map of the FM, (b) the 2D cortical map of the CR, and (c) the 2D cortical map of the mean-based algorithm. (d) The functional median log-periodogram of each channel (FM), (e) the functional median log-periodogram of each channel (CR), and (f) the functional mean log-periodograms of each channel.}
\label{last}
\end{figure}

We apply the three clustering algorithms to the data from the first and the last minutes of the low-frequency band. We show the results by 2D cortical maps in Figure \ref{first} and Figure \ref{last}. 
When applying FM and CR, we will show the functional median in each channel (same color in one cluster), and we will show the functional means of the channels when applying the mean-based algorithm.

As shown in Figure \ref{first} and Figure \ref{last}, in summary, we have the following findings:
\begin{enumerate}

\item All three algorithms give us approximately symmetric clustering results in terms of left-right side (CR algorithm is not so obvious). FM and the mean-based algorithms give similar clusters with more clear margins, but they tend to cluster a large number of channels (black) together while several channels are isolated as single-member clusters. In contrast, the number of channels in each cluster is evenly distributed in CR. Also, on the back side of the scalp, FM and mean-based cluster all the channels together while the central region based algorithm assign these channels to different clusters.
\item The first and the last minutes represent a resting-state test, and there is no significant difference between their clustering results.
\item The functional median (functional mean) log-periodograms of each channel show that anterior (front) and posterior (back) channels have a higher power than others on the low-frequency band.

\end{enumerate}

We compare the clustering results of the EEG data for the same subject  to \cite{euan2018spectral} (shown as Figure 18(b)), we found that:
\begin{enumerate}

\item In the anterior (front) region of the brain, \cite{euan2018spectral} gives one cluster while we give multiple clusters. 

\item In the posterior (back) region of the brain, \cite{euan2018spectral} gives multiple clusters, which is similar to CR but differs from the FM and mean-based algorithms.

\item In both \cite{euan2018spectral} and our algorithms, there is no significant difference between the early and late stage.

\end{enumerate}

In the supplementary material, we provide the rank sum test \citep{lopez2009concept} of the identified clusters and the clustering results of another subject in Wu et al. (2014).

\subsection{Robustness of the Clustering Algorithms}
To test the robustness and the stability of the clustering algorithm, we considered $15$ subsets of the 3-minute EEG recordings by applying a moving window with a length of $30$ seconds and moving forward $10$ seconds per step. Since the EEG data we have analyzed are during the resting state, we expect the $15$ clustering results are similar for the entire $3$ minutes.

\begin{figure}
\centering
\raisebox{-2cm}{\includegraphics[width=\textwidth]{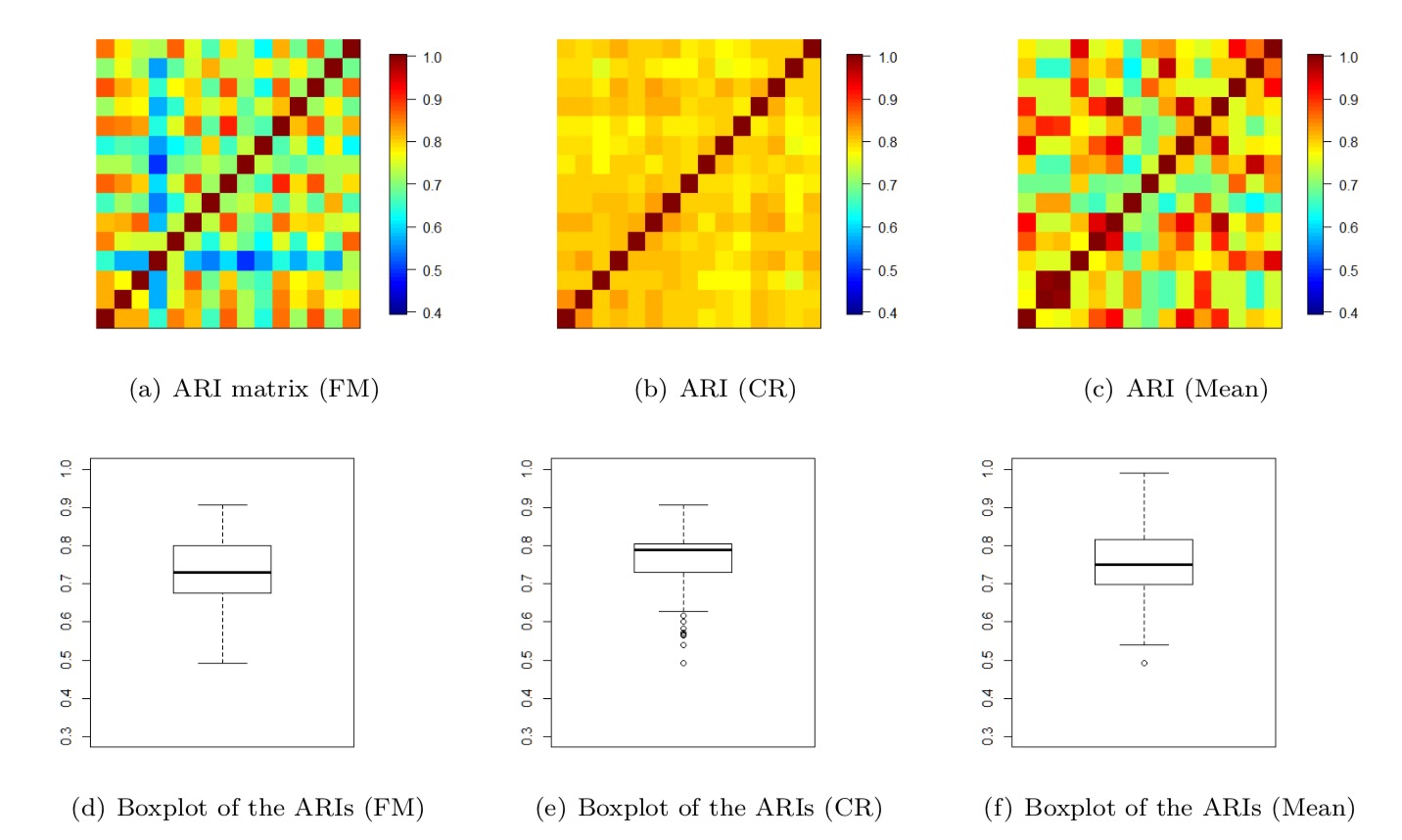}}

\linespread{0.3}
\caption{(a),(b),(c) The RI matrix of the three algorithms, the RI in the $i$-th row and $j$-th column measures the similarity of the clustering results from the $i$-th and $j$-th windows ($i,j=1,...,15$).  (d), (e), (f) The boxplots of the $105$ ($\frac{15\times 14}{2}$) RIs in each of the three algorithms.}
\label{rand}
\end{figure}

We applied the three clustering algorithms to the 15 datasets with the same number of clusters, and compute the ARI values for any two of the $15$ clustering results. To make the test fair, we set fixed the number of clusters as $7$ Figure \ref{rand} shows the resulting $15$ by $15$ ARI values for each clustering algorithms and the corresponding boxplots of all the off-diagnose ARI values.

Since higher values of the ARI with smaller variability imply more consistent clustering results, we can see that the CR based algorithm gives the most robust and stable clustering during the $3$ minutes, while the median and the mean based algorithms are similar regarding the consistency. It is not surprising because the CR based algorithm considers the majority of the epochs for clustering while the FM based algorithm only uses the median, and the mean-based algorithm is not resistant to outliers.

As shown in Figure \ref{rand}, we can see that CR have the best performance: it has highest similarities in general with the lowest variability. FM and mean based algorithm have similar performance, they have similar medians and large variability. One reason to explain this is that FM only considers the most central epoch in each cluster, mean based algorithm are easily influenced by the contaminations, while CR considers the information of half of the epochs, and is robust to contaminations.

\section{Conclusion and Discussion}
In this paper, we proposed two clustering algorithms for functional data with replicates, using a frequency domain approach. We introduced the functional data ranking methodology in the functional data analysis to study spectral densities for time series data. The proposed methods are suitable for clustering functional data with replicates. In the hierarchical merging algorithms, we proposed two new dissimilarity measures, the distance between functional medians, and the size of the central region when merging, both of which are rank-based and thus robust in the presence of outliers. In the simulation studies, we have shown that the proposed clustering algorithms are particularly attractive when data are noisy. For the settings we have considered in EEG data application, the CR algorithm shows the best performance in robustness because it accounts for the variability of the majority, or the $50\%$ most representative curves, rather than choosing only one median curve. One potential limitation of the proposed clustering algorithms is the computational cost. Specifically, at each merging step, computing the functional median or the central region requires more computations.  One extension of this paper in the future is to compare the resting-state EEG data with stimuli by using the clustering algorithm that we proposed. It is also worth pointing out that the functional ranking methodology allows us to test the significance of the identified clusters as illustrated in the EEG application.

  \section{Acknowledgements}

The authors thank Professor Wu for sharing the EEG data set. The authors would also
like to thank the Editor and the anonymous Associate Editor for their
suggestions.

\bibliography{a1}

\end{document}